\documentclass[fp,twocolumn]{jpsj3}
\usepackage{txfonts}
\usepackage[dvipdfmx]{graphicx}				
\usepackage{physics}
\usepackage{ulem}							
\usepackage{amssymb}
\usepackage{bm,color}

\title{Spin-flip Scattering at a Chiral Interface of Helical Chains}

\author{Keita Matsubara and Kazumasa Hattori}
\inst{Department of physics, Tokyo Metropolitan University, 1-1 Minami-Osawa, Hachioji, Tokyo 192-0397, Japan} 

\abst{We investigate spin-flip scattering processes of electrons when they pass a chiral interface, which is the boundary between right- and left-handed chiral structure. We construct a minimal $p$-orbital model consisting of the right- and left-handed one-dimensional threefold helical chains connected at $z=0$ with the nearest neighbor hopping and the spin-orbit coupling. The dynamics of a spin-polarized wave packet passing through the interface, the Green's functions, and electronic states near the interface are analyzed numerically. We find that the microscopic structure of the interface is important and this strongly affects the local electronic orbital state. This in addition to the spin-orbit coupling determines whether the spin flip occurs or not at the chiral interface and suggests a possible spin transport control by the orbital configuration at the chiral interface.
}

\kword{chiral crystal, interface, spin flip}

\begin{document}
\maketitle

\section{Introduction}
Chiral crystals lack mirror and space inversion symmetries and possess the two types of structures distinguished by handedness or chirality\cite{kelvin}. The two structures are the mirror images with each other and called right-handed (RH) and left-handed (LH). In recent years, physics in chiral crystals have attracted great attention due to their unique properties\cite{def_of_chirality}. For example, the absence of the mirror and the space inversion symmetry leads to cross-correlated responses\cite{cross_correlation} such as Edelstein effect\cite{edelstein,CIM_Te} and electorotation effects.\cite{chiral_rot} Phonons in chiral crystals, chiral phonons, \cite{obs_chiral_phonons,chiral_phonons,truly} have their own angular momenta locked by the direction of their momentum and it is known that their dispersions show the band splitting for different angular momenta.\cite{cp,cp_disp}  The angular momentum of phonons gives correction to the Einstein-de-Haas effect\cite{CAM_EdH} and it can also couple with other quantities such as electrons' spin. \cite{chiral_phonon_spin,magnon-phonon} 

Among various phenomena related to the chirality, spin-dependent and/or non-reciprocal spin transports are the key to the deeper understanding of the physics behind the chirality degree of freedom. Anisotropic transports in chiral materials such as carbon nanotubes\cite{c_nano_review} and in chiral magnets\cite{MnSi,meso,spin_curent_chiral_mag}  and the transport of the ``spin'' angular momentum of phonons at interfaces\cite{chiral_phonon_interface} have been actively studied over the past two decades. Moreover, spin selective transports in chiral molecules called chirality-induced spin selectivity (CISS), finite spin polarizations of electrons passing through chiral materials, have also been extensively studied.\cite{CISS,CISS_review} 

To realize the ``full'' responses characteristic to the chirality in the chiral systems, controlling their domains is crucially important. So far, there is no established method to selectively synthesize mono chiral domain for general chiral materials. Recent studies on the basis of multipole theory\cite{chiral_rot,Inda,flo} show the importance of electric toroidal monopole in chirality enantioselection. At the boundary between the LH and RH crystals, a unique interface is formed and we call it chiral interface (CIF). Such CIFs inevitably exist in chiral crystals, and thus, one can regard them as microscopic chirality ``junctions'' in bulk systems. 
In chiral magnets, such structures lead to large magnetoresistance\cite{CrNb3S6} and investigated theoretically\cite{magnetotrans,spiral_magnet}. Spin-dependent phenomena in magnetic junction systems such as the tunneling magnetoresistance\cite{TMR_review} and the spin-to-charge conversion\cite{SCC} have also attracted great attention and play one of primal roles in recent nanotechnological developments. The focused-ion beam technology\cite{FIB} also allows one to observe a single domain wall (if it exists) dynamics and anisotropic transport in artificially constructed structure.  In this respect, it is highly important to understand the spin transport physics at these CIFs.

\begin{figure}[t]
\centering
\includegraphics[width=1.0\linewidth]{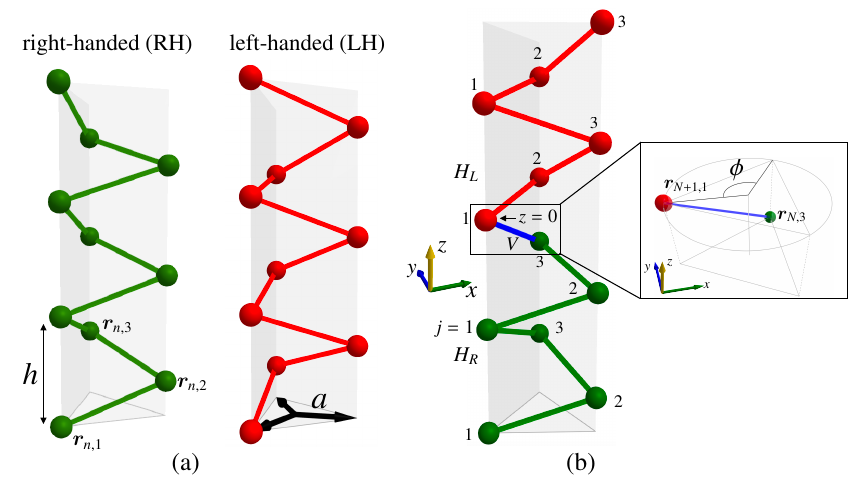}
\caption{(Color online) (a) The right-handed (RH; green) and left-handed (LH; red) chain. Each unit cell contains three sublattices indexed as $\bm{r}_{n,j}$ ($j = 1,2,3$). (b) One-dimensional chiral chain with the CIF. The blue line represents the inter-chain bond. The inset shows the parameter $\phi$. }
\label{chains}
\end{figure}

In this paper, we investigate how the CIF between RH and LH systems affects the spin polarization of electrons passing through the CIF in chiral crystals. We analyze a minimal one-dimensional $p$-orbital tight-binding model with RH and LH  three-fold helices connected at the origin $(z=0)$. See Fig.~\ref{chains}. We clarify the dynamics of a spin-polarized electron wave packet passing through the CIF and the mechanism of spin-flip processes at the CIF. 

This paper is organized as follows.  We will introduce our model Hamiltonian in Sec.~\ref{sec:model} and introduce our methods used in the analysis in Sec.~\ref{sec:method}. In Sec.~\ref{sec:results}, we will show the numerical results of CIF scattering and analyze the spin-flip processes at the CIF. Section \ref{sec:discussion} is devoted to the discussion about the mechanism of the spin flip at the CIF, focusing on the localized modes at the CIF and their orbital character. Finally, we summarize the results of this paper in Sec.~\ref{sec:conclusion}.
\section{Model}\label{sec:model}
In this section, we provide our model Hamiltonian. We consider a one-dimensional three-fold helical chain with a CIF at its center. Such a three-fold helical structure is, e.g., found in Tellurium.\cite{CIM_Te} Our model is a simplified one taking a single helical chain in the Te-like crystal structure, and one of minimal models for analyzing the electron scattering at the CIF. Instead of taking into account detail material specific parameters, we try to understand the microscopic mechanism of spin-flip scatterings at CIFs by analyzing a minimal model introduced in this Sec.~\ref{modelH}.

\subsection{Model Hamiltonian}\label{modelH}
The system we will analyze in this paper consists of two three-fold helical chains along the $z$-direction with different handedness as shown in  Fig.~\ref{chains}(a) and we connect them at $z=0$. This generates a CIF [Fig.~\ref{chains}(b)]. We assume that the half chain for $z<0$ is the RH one and label the site position $\bm{r}_{n,j}=(-a\cos\theta_j,-a\sin\theta_j,z_{n,j})$ with $\theta_j = \frac{2\pi}{3}(j-1)$ and $z_{n,j} = h[n-N-1+\frac{h}{3}(j-1)]$. Here, $n$ $(= 1,2,\cdots, N)$ is the index for the unit cell in the RH chain and $j$ $(=1,2,3)$ is that for the sublattice. Similarly, the $z\geq0$ part is the LH one with the site position  $\bm{r}_{n,j} = [-a{\rm cos}(\theta_{j+1}-\phi),\ a{\rm sin}(\theta_{j+1}-\phi),\ z_{n,j}]$ for $n = N+1,N+2,\cdots, 2N$, where we introduce the parameter $\phi$ as shown in the inset of Fig.~\ref{chains}(b). This determines the structure of the CIF at $z = z_{\text{CIF}}\equiv(z_{N,3} + z_{N+1,1})/2$. Hereafter, we set $a =1$ and $h=1$.

Next, we explain the Hamiltonian used in this study. To discuss the correlation between the chirality and spins, one needs to consider the spin-orbit coupling (SOC). One of the minimal setting is a $p$-electron system with three $p_x,p_y$ and $p_z$ orbitals since they have finite SOC. Recently, multi-center spin-orbit interactions, i.e., spin-dependent hoppings, have been investigated for helical structure\cite{SOC_micro}, but here such terms are ignored as a zeroth approximation. The orbital degrees of freedom can take into account the effects of helical structure. We use a vector form $\hat{\bm{p}}^\dagger_{n,j,\sigma} = (\hat{p}^\dagger_{n,j,x,\sigma},\hat{p}^\dagger_{n,j,y,\sigma},\hat{p}^\dagger_{n,j,z,\sigma})$, where $\hat{p}^\dagger_{n,j,\ell,\sigma}$ is the creation operator of $p_\ell$ ($\ell=x,y,z$) electron at $\bm{r}_{n,j}$ with the spin $\sigma (= \uparrow,\downarrow)$. For simplicity, we have ignored the orbital dependent local crystal field. Hamiltonian consists of the intra-chain part $\hat{H}_{R,L}$ and the inter-chain part $\hat{V}$ as $\hat{H} = \hat{H}_{R} + \hat{H}_{L} + \hat{V}$, where
\begin{align}
&\hat{H}_{R (L)}=\sum_{n,i,\sigma}^{\substack{1\leq n \leq N \\(N+1\leq n\leq 2N)}}(\hat{\bm{p}}^\dagger_{n,j+1,\sigma}\cdot\bm{d}^n_{j,j+1})(\hat{\bm{p}}_{n,j,\sigma}\cdot\bm{d}^n_{j,j+1})+\text{h.c.}\notag\\
&\qquad\qquad+\lambda\sum_{n,j,\ell,\ell',\sigma,\sigma'}^{\substack{1\leq n \leq N \\(N+1\leq n\leq 2N)}}\hat{p}^\dagger_{n,j,\ell,\sigma}(\bm{L}\cdot\bm{S})_{p_{\ell,\sigma},p_{\ell',\sigma'}}\hat{p}_{n,j,\ell',\sigma'}\label{HR},\\
&\hat{V}=  \hat{V}_{LR} + \hat{V}_{RL}\label{V_inter},\\
&\hat{V}_{LR}= \sum_{\sigma}(\hat{\bm{p}}^\dagger_{N+1,1,\sigma}\cdot\bm{d}_{\text{CIF}})(\hat{\bm{p}}_{N,3,\sigma}\cdot\bm{d}_{\text{CIF}}) = \hat{V}^\dagger_{RL}.
\label{full}
\end{align}
Here, the first term in the RHS of Eq.~(\ref{HR}) describes the nearest neighbor (NN) hopping and the second one does the SOC with the coupling constant $\lambda$. $\hat{V}_{LR(RL)}$ in Eq.~(\ref{full}) represents the hopping from RH (LH) to the LH (RH) chain at the CIF. We have set the NN hopping to 1 as a unit of energy and $\bm{d}^n_{j,j+1} = ( \bm{r}_{n,j+1}-\bm{r}_{n,j})/|\bm{r}_{n,j+1}-\bm{r}_{n,j}|$, where $\bm{r}_{n,4}$ should be regarded as $\bm{r}_{n+1,1}$. We take into account only the $\sigma$-bond\cite{sk_para} for the NN hoppings, and thus, the electrons in the orbitals along the bond direction $\bm{d}_{j,j+1}^n$ can hop, while the others cannot.  Although $\pi$-bond {hoppings} exist in general, those for $\sigma$-bond  are usually larger as also seen in Te\cite{trigonal_Te}. Thus, they are ignored here. A model taking into account $\pi$-bond hoppings is shown in Appendix A and their effects are briefly analyzed. $\bm{d}_{\text{CIF}} = (\bm{r}_{N+1,1}-\bm{r}_{N,3})/|\bm{r}_{N+1,1}-\bm{r}_{N,3}|$ is the bond direction at the CIF and depends on $\phi$. Although variations in $\phi$ cause changes in the bond length, we set the absolute value of the hopping unchanged. This is because it is not important in comparison to the change in the direction. $(\bm{L}\cdot\bm{S})_{p_{\ell,\sigma},p_{\ell'\sigma'}}$ is the $(p_{\ell,\sigma},p_{\ell'\sigma'})$ component of the matrix $\bm{L}\cdot\bm{S}$. This has a form in the basis of $\lbrace p_{x\uparrow},p_{x\downarrow},p_{y\uparrow},p_{y\downarrow},p_{z\uparrow},p_{z\downarrow}\rbrace$ as
\begin{align}
\bm{L}\cdot\bm{S} = \frac{1}{2}
\begin{pmatrix}
0&0&i&0&0&1\\
0&0&0&-i&-1&0\\
-i&0&0&0&0&i\\
0&i&0&0&i&0\\
0&-1&0&-i&0&0\\
1&0&-i&0&0&0\\
\end{pmatrix},
\label{SOC_mat}
\end{align}
where $\bm{L} = (L_x,L_y,L_z)$ and $\bm{S} = (S_x,S_y,S_z)$ are the orbital and spin angular momentum matrix, respectively. 

\subsection{The eigenenergy and the eigenstates}
$\hat{H}$ [Eqs.~(\ref{HR})-(\ref{full})] can be easily diagonalized numerically with the open boundary condition at $\bm{r}_{1,1}$ and $\bm{r}_{2N,3}$ and one obtains the eigenenergy $\varepsilon_m$ as 
\begin{align}
\hat{H} = \sum_{m=1}^{36N}\varepsilon_m\hat{c}^\dagger_m\hat{c}_m.\label{eigen_full}
\end{align}
Here,  $\hat{c}^\dagger_m$ is the creation operator of the $m$th eigenstate $\ket{m}=\hat{c}^\dagger_m\ket{0}$, where $\ket{0}$ is the vacuum. $\hat{c}^\dagger_m$ can be written as a linear combination of $\hat{p}^\dagger_{n,j,\ell,\sigma}$ as 
\begin{align}
	\hat{c}^\dagger_m = \sum_{n,j,\ell,\sigma}T_{n,j,\ell,\sigma;m}\hat{p}^\dagger_{n,j,\ell,\sigma},\quad 
	\hat{p}^\dagger_{n,j,\ell,\sigma} = \sum_mT^*_{n,j,\ell,\sigma;m}\hat{c}^\dagger_m, \label{eq:p_c_rel}
\end{align}	
where $T_{n,j,\ell,\sigma;m}$ is a $36N\times36N$ unitary matrix. Since the time-reversal symmetry is present, $\varepsilon_m$ is at least two-fold degenerate. Note that there is no translational symmetry and thus the wavenumbers are not good quantum numbers.
\section{Methods} \label{sec:method}
Here, we explain methods for our analyses of electron scatterings at the CIF. First, we consider a situation where a spin polarized electron is injected to the RH chain at the initial time $t=0$. The wave packet diffuses as the time passes. It reaches the CIF and then is scattered. We derive an expression for the dynamics of the wave packet propagating on the filled Fermi sea in contrast to one-particle wave packet studies, for example, in mesoscopic ring devises \cite{ring} and a monolayer graphene\cite{mono,massbari}. We discuss its spin, magnetization, and the probability weight of Bloch states constituting the wave packet in each of the RH and LH chain. We also carry out analyses based on the Green's functions in order to investigate the scattering at the CIF microscopically. All the following analysis is carried out at the temperature $T=0$ and we set $\hbar = 1$ throughout this paper.
\subsection{Time dependence of the observables}\label{section_time}
Let us first define the initial state $\ket{I}$ :  an electron with the spin $\sigma_I$ is added at $\bm{r}_{n_I,j_I}$ in the $p_{\ell_I}$-orbital to the ground state $\ket{F}$ at $t=0$, which reads as
\begin{align}
\ket{I}&\equiv\frac{1}{\sqrt{A}}\hat{p}^\dagger_{n_I,j_I,\ell_I,\sigma_I}\ket{F},\quad \Big(\ket{F} \equiv \prod_{m,\varepsilon_m \leq \varepsilon_F}\hat{c}^\dagger_m\ket{0}\Big),
\end{align}
where $\ket{0}$ is the vacuum, $\varepsilon_F$ is the Fermi energy, and $A$ is the normalization constant defined as $A = \sum_{m,\varepsilon_m>\varepsilon_F}|T_{n_I,j_I,\ell_I,\sigma_I;m}|^2$. In the Heisenberg picture, the $t$ dependence of the expectation value of operator $\hat{O}(\bm{r}_{n,j},t)\equiv \sum_{\ell,\ell',\sigma,\sigma'}\hat{p}^\dagger_{n,j,\ell,\sigma}(t)(O)_{p_{\ell,\sigma},p_{\ell'\sigma'}}\hat{p}_{n,j,\ell',\sigma'}(t)$ can be calculated as 
\begin{align}
&O(\bm{r}_{n,j},t)\equiv\ev{\hat{O}(\bm{r}_{n,j},t)}{I}\notag\\
 =&\sum_{\substack{m,m'\\\ell,\ell'\sigma,\sigma'}}T^*_{n,j,\ell,\sigma;m}(O)_{p_{\ell,\sigma},p_{\ell'\sigma'}}T_{n,j,\ell',\sigma';m'}e^{i(\varepsilon_m-\varepsilon_{m'})t }\ev{\hat{c}^\dagger_m\hat{c}_{m'}}_{I},
\label{Sz_time}
\end{align}
 where $\ev{\cdots}_{I} \equiv \ev{\cdots}{I}$ and we have used the relation Eq.~(\ref{eq:p_c_rel}) and $\hat{c}^\dagger_m(t) = e^{i\varepsilon_mt}\hat{c}^\dagger_m$. The expectation value of the spin $S_{\nu}(\bm{r}_{n,j},t)$ ($\nu = x,y,z$) and the magnetization $M_{\nu}(\bm{r}_{n,j},t)$ can be obtained by replacing $O$ with the spin matrix $S_\nu$ or the magnetization matrix $M_{\nu} = 2S_{\nu} + L_{\nu}$. We note that the presence of the filled Fermi sea prevents the injected electrons from being scattered into the eigenstates with $\varepsilon_m<\varepsilon_F$. For one-particle description of wavepacket, such scattering processes are allowed at the interface\cite{massbari}. In this sense our formulation takes into account many-particle effects even in non-interacting  systems. 
 
 \subsection{The Bloch states constituting the wave packet}
We now analyze the Bloch states $\ket{R(L);k,\mu}$ constituting the wave packet scattered at the CIF. Although the wavenumbers are not good quantum numbers, we can project the eigenstate $\ket{m}$ to the Bloch state $\ket{R(L);k,\mu}$ with the wavenumber $k$ and the band index $\mu\;(=1,2,\cdots,18)$ in the RH-only (LH-only) system. Here, we apply the periodic boundary condition (PBC) to obtain $\ket{R(L);k,\mu}$ and the wavenumber $k$ is a good quantum number in this system. Thus, we can obtain the weight of the Bloch states in the wave packet for any $t$. The Hamiltonian in the RH- or LH-only system under the PBC is diagonalized as
 \begin{align}
 H_{R} = \sum_{k,\mu}E^R_{k,\mu}\hat{R}^\dagger_{k,\mu}\hat{R}_{k,\mu},\;H_{L} =  \sum_{k,\mu}E^L_{k,\mu}\hat{L}^\dagger_{k,\mu}\hat{L}_{k,\mu},\label{single-handed}
 \end{align}
where $E^R_{k,\mu}(E^L_{k,\mu})$ and $\hat{R}^\dagger_{k,\mu}(\hat{L}^\dagger_{k,\mu})$ are the eigen energy and the creation operator of the Bloch state $\ket{R(L);k,\mu}$, respectively. $\hat{R}^\dagger_{k,\mu}(\hat{L}^\dagger_{k,\mu})$ is defined as $\hat{R}^\dagger_{k,\mu}(\hat{L}^\dagger_{k,\mu})\equiv \sum_{n,j,\ell,\sigma}B^{R(L)}_{j,\ell,\sigma;\mu}(k)\hat{p}^\dagger_{n,j,\ell,\sigma}$. Here, $B^{R(L)}_{j,\ell,\sigma;\mu}(k)$ is the $18\times18$ matrix for each $k$, where the matrix elements are labeled by $\lbrace j,\ell,\sigma\rbrace$ and $\mu$.

$\hat{c}^\dagger_m(t)$ in Eq.~(\ref{eigen_full}) can be represented by $\hat{R}^\dagger_{k,\mu}(t)[\hat{L}^\dagger_{k,\mu}(t)]$ as
\begin{align}
\hat{c}^\dagger_m(t) = \sum_{k,\mu}\big[U^{R*}_{k,\mu;m}\hat{R}^\dagger_{k,\mu}(t) +U^{L*}_{k,\mu;m}\hat{L}^\dagger_{k,\mu}(t) \big],
\label{eigen_wave}
\end{align}
where $U^{R(L)*}_{k,\mu;m} = \bra{R(L)k,\mu}\ket{m}$. This allows us to calculate the probability of finding the Bloch state $\ket{R(L);k,\mu}$ in the RH (LH) chain $w^R_{k,\mu}(t)$ [$w^L_{k,\mu}(t)$] as
\begin{align}
w^R_{k,\mu}(t) = \ev{\hat{R}^\dagger_{k,\mu}(t)\hat{R}_{k,\mu}(t)}{I},\;w^L_{k,\mu}(t) = \ev{\hat{L}^\dagger_{k,\mu}(t)\hat{L}_{k,\mu}(t)}{I}.\label{weight}
\end{align}
When the wave packet stays in the RH chain for small $t$, we obtain $w^R_{k,\mu}(t) \ne0$ for some $\ket{R;k,\mu}$ with $E^R_{k,\mu}\gtrsim\varepsilon_F$ and $w^L_{k,\mu}(t) =0$ for any $\ket{L;k,\mu}$ with $E^L_{k,\mu}\gtrsim\varepsilon_F$. Here, we assume the initial position $n_I$ is sufficiently far from the CIF and the approximate equality is due to the finite size effects. However, as soon as the wave packet enters the LH chain, $w^L_{k,\mu}(t)$ also becomes finite for some $\ket{L;k,\mu}$ with $E^L_{k,\mu}\gtrsim\varepsilon_F$, which reflects the scattering profile of the wave packet at the CIF. The bands with the energy $E^{R(L)}_{k,\mu}\lesssim\varepsilon_F$ are assumed to be filled and do not change with $t$, which is valid at $T=0$.
\subsection{Scattering Green's function}
\label{green_method}
As an alternative approach, we here consider one-electron scattering problems. The information of the microscopic one-electron processes is useful for the analysis of the scattering at the CIF discussed in Sec. \ref{section_time}. We calculate the probability of finding a Bloch state $\ket{L;k,\mu}$ in the LH chain at $t>0$ with initial Bloch state $\ket{R;k',\mu'}$ at $t=0$ in the RH chain. Since the wave packet considered in Sec.~\ref{section_time} consists of multiple Bloch states, information of the scattering of a specific Bloch state is important as their  elementary information. 

The electron is injected in the RH chain at initial time $t=0$ and then scattered at the CIF as discussed in Sec.~\ref{section_time}. However, this time the injected electron is in the Bloch state $\ket{R;k',\mu'} = \hat{R}^\dagger_{k',\mu'}\ket{0}$. We denote the probability of finding the electron in the Bloch state $\ket{L;k,\mu} = \hat{L}^\dagger_{k,\mu}\ket{0}$ in the LH chain for $t>0$ as $|P_{k,\mu,k',\mu'}(t)|$, where
\begin{align}
P_{k,\mu,k',\mu'}(t)&\equiv\ev{\hat{L}_{k,\mu}(t)\hat{R}^\dagger_{k',\mu'}(0)}{F}\theta(t), \label{TP_def}
\end{align}
is a kind of retarded Green's function. Here, $\theta(t)$ is the Heaviside's step function. Using Eq.~(\ref{eigen_wave}) and the Fourier transformation, one obtains
\begin{align}
P_{k,\mu,k',\mu'}(\omega) = \sum_{\varepsilon_m > \varepsilon_F}\frac{U^{L*}_{k,\mu;m}U_{k',\mu';m}^{R}}{\omega-\varepsilon_m + i\delta} \quad\text{ with $0<\delta\ll1$}.
\label{trans_prob}
\end{align}
Here, $\omega$ is the frequency (energy). $|P_{k,\mu,k',\mu'}(\omega)|^2$ gives the scattering probability of finding an electron with the wavenumber $k$ and the energy $\omega$ in the LH chain when the electron is injected to the RH chain with the wavenumber $k'$ and the energy $E_{k',\mu'}^R$.
\section{Results} \label{sec:results}
\begin{figure}
\centering
\includegraphics[width=1.0\linewidth]{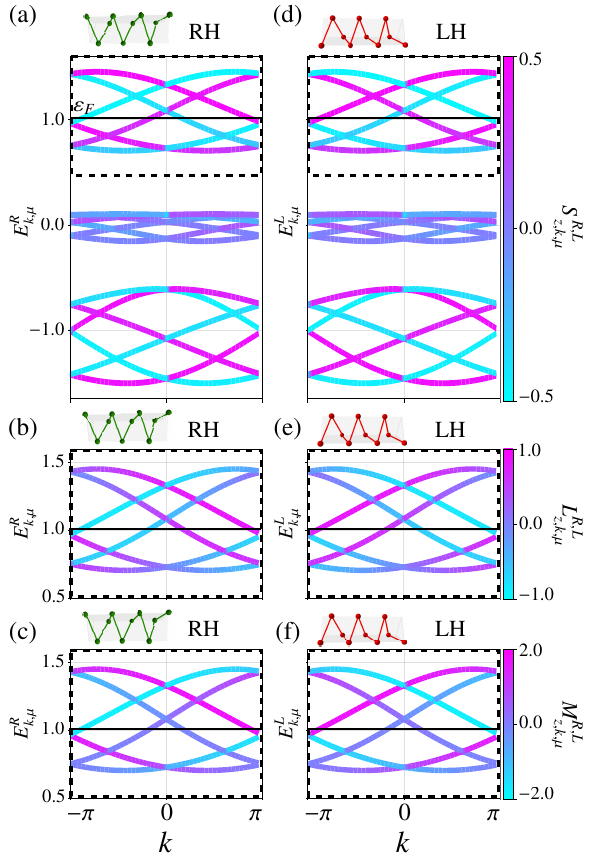}
\caption{(Color online) Band dispersion of RH chain [(a)-(c)] and LH chain [(d)-(f)]. The color represents the expectation value of the spin angular momentum for the Bloch state $S_{z,k,\mu}^{R(L)}$ in (a) and (d), the orbital angular momentum $L_{z,k,\mu}^{R(L)}$ in (b) and (e) and magnetization $M_{z,k,\mu}^{R(L)} = L_{z,k,\mu}^{R(L)} + 2S_{z,k,\mu}^{R(L)}$ in (c) and (f). For (b), (c), (e), and (f), the upper part of the dispersions enclosed by the dashed line in (a) and (d) is shown.
}
\label{band_spin}
\end{figure}

In this section, we show our numerical results. To introduce the basic single-electron properties in the RH (LH) systems, we first discuss the dispersion of the single-handed chain $E^{R(L)}_{k,\mu}$ in Eq.~(\ref{single-handed}), the spin and the orbital angular momentum, and the magnetization for each band. Secondly, we demonstrate the time evolution of spin angular momentum and magnetization when an electron is injected in the RH chain at $t=0$ on the basis of Eq.~(\ref{Sz_time}). We also examine how the Bloch states forming the wave packets are scattered at the CIF. Then, we calculate the profile of $|P_{k,\mu,k',\mu'}(\omega)|^2$ [Eq.~(\ref{trans_prob})] and examine whether the CIF structure, i.e., $\phi$ dependence affects the profile of $|P_{k,\mu,k',\mu'}(\omega)|^2$. Lastly, we show the density of states (DOS) and the wave functions to analyze the spin-flip scattering in detail. Throughout this section, we set $N=50$, $\lambda = 0.5$ and the filling $n_f$ is $n_f=5$ per unit cell, which leads to $\varepsilon_F\simeq 1.01$. Thus, we can focus on the upper six bands as enclosed by the dashed line in Fig.~\ref{band_spin}. We have checked that the results for $N=50$ is qualitatively unchanged even for larger system sizes up to $N=100$.

\subsection{Band dispersion}
Figure~\ref{band_spin} shows the band dispersion of the single-handed chain. The thick horizontal line represents the Fermi energy $\varepsilon_F$. The color represents the expectation value of the spin angular momentum $S_{z,k,\mu}^{R(L)}\equiv\sum_{i=1,2,3}{}_i\!\ev{S_z}{R(L);k,\mu}_i$ [(a) and (d)], the orbital angular momentum $L_{z,k,\mu}^{R(L)}\equiv\sum_{i=1,2,3}{}_i\!\ev{L_z}{R(L);k,\mu}_i$ [(b) and (e)], and the magnetization $M_{z,k,\mu}^{R(L)} = L_{z,k,\mu}^{R(L)} + 2S_{z,k,\mu}^{R(L)}$ [(c) and (f)], where $\ket{R(L);k,\mu}_i$ is the $i$th sublattice component of the Bloch state. For Figs.~\ref{band_spin}(b), \ref{band_spin}(c), \ref{band_spin}(e), and \ref{band_spin}(f), only the upper six bands are shown since the other bands below the Fermi energy are already filled and are not important in the analyses in the following. It is clear that the Bloch state $\ket{R(L);k,\mu}$ is non degenerate owing to the SOC and the lack of inversion symmetry except for the high-symmetry or accidental points. Note that $S_{z,k,\mu}^{R} = -S_{z,-k,\mu}^{R} = -S_{z,k,\mu}^{L}= S_{z,-k,\mu}^{L}$ and the similar relations hold for  $L_{z,k,\mu}^{R(L)}$ and $M_{z,k,\mu}^{R(L)}$. This is due to the time reversal symmetry and the lack of the $z$ mirror symmetry.
\begin{figure}[h]
\centering
\includegraphics[width=0.9\linewidth]{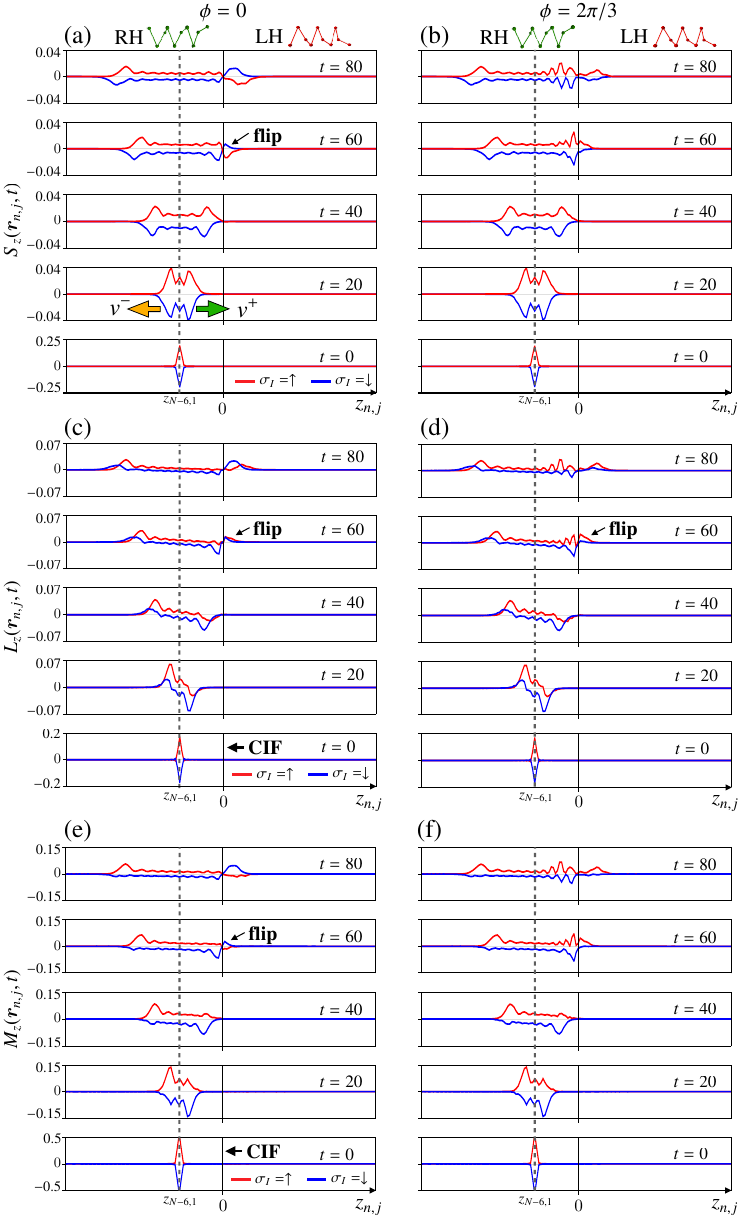}
\caption{(Color online) $S_z(\bm{r}_{n,j},t)$ with $t=0,20,40,60$, and 80 for (a) $\phi=0$ and (b) $\phi=2\pi/3$, $L_z(\bm{r}_{n,j},t)$ for (c) $\phi=0$ and (d) $\phi=2\pi/3$, and $M_z(\bm{r}_{n,j},t)$ for (e) $\phi=0$ and (f) $\phi=2\pi/3$. The data for $\sigma_I=\downarrow(\uparrow)$ are colored in red (blue) and the initial position is indicated by the vertical dashed line.}
\label{Sz_snap}
\end{figure}
\begin{figure}[h]
\centering
\includegraphics[width=0.9\linewidth]{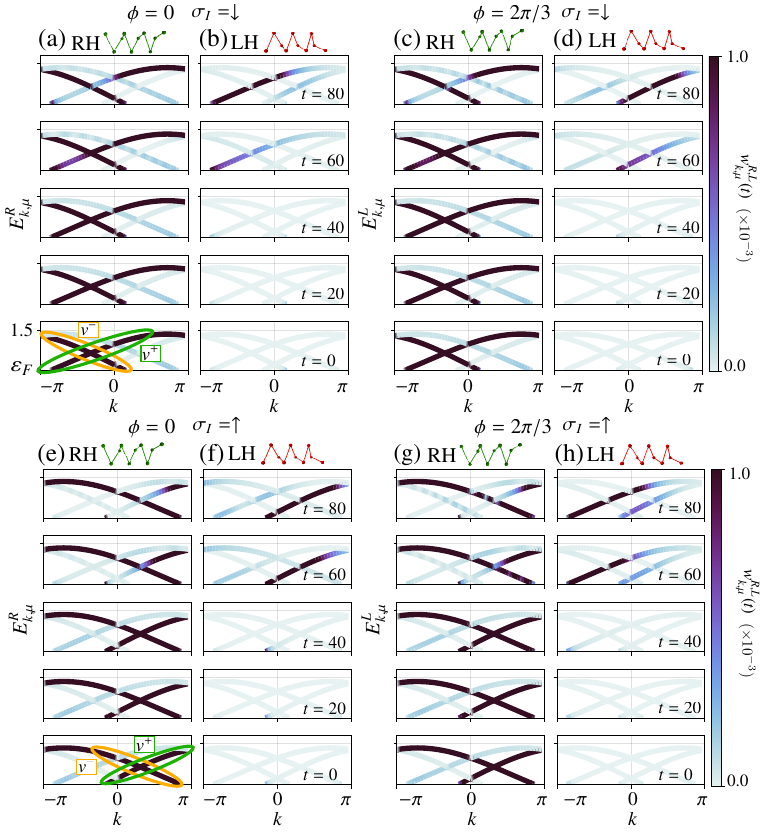}
\caption{(Color online) The time evolution of the weight of the Bloch states constituting the wave packet $w^{R(L)}_{k,\mu}(t)$ for $\phi=0$ and $\sigma_I=\downarrow$ [(a) for RH and (b) for LH], for $\phi=2\pi/3$ and $\sigma_I=\downarrow$ [(c) for RH and (d) for LH], for $\phi=0$ and $\sigma=\uparrow$ [(e) for RH and (f) for LH], and for $\phi=2\pi/3$ and $\sigma_I=\uparrow$ [(g) for RH and (h) for LH].}
\label{weight_snap}
\end{figure}

\subsection{Time evolution of spin and magnetization}\label{subsec_timeevo}
Figure~\ref{Sz_snap} illustrates the time evolution of the spin density $S_z(\bm{r}_{n,j},t)$ for (a) $\phi=0$ and (b) $\phi = 2\pi/3$, the orbital angular momentum $L_z(\bm{r}_{n,j},t)$ for (c) $\phi=0$ and (d) $\phi = 2\pi/3$, and the magnetization $M_z(\bm{r}_{n,j},t)$ for (e) $\phi=0$ and (f) $\phi=2\pi/3$. See Eq.~(\ref{Sz_time}). Here, the unit of time is $\frac{1}{[\text{NN hopping in eV}]}\times10^{-15}$ s. At $t=0$, a spin up ($\sigma_I=\uparrow$; red line) or down ($\sigma_I=\downarrow$; blue line) electron is injected to the RH chain at $(n_I,j_I,\ell_I) = (N-6,1,x) $,  which results in the nearly delta function-like peak at $z_{N-6,1}$ at $t=0$. This particular choice of initial state does not qualitatively affect the final results. We have chosen $\ell_I = x$ since it creates clearer wave packet than $\ell_I = y$ and $\ell_I = z$, and thus, it is useful for the analyses. As the time evolves, the wave packet diffuses into a wave with positive group velocity ($v^+$-wave propagating in the positive $z$ direction) and with negative group velocity ($v^-$-wave propagating in the negative $z$ direction).  We concentrate on the former in the following. When $t\sim60$, the $v^+$ wave reaches the CIF and the spin flip is observed for $\phi=0$, while not for $\phi=2\pi/3$. 

Figures~\ref{weight_snap}(a) and \ref{weight_snap}(b) show the $k$-resolved weight of the wave packets $w^{R,L}_{k,\mu}(t)$ for $\sigma_I = \downarrow$ [Eq.~(\ref{weight})] in the RH and LH chain, respectively for $\phi=0$. Similarly, Figs.~\ref{weight_snap}(c) and \ref{weight_snap}(d) show that for $\phi=2\pi/3$. For $\sigma_I=\uparrow$, the corresponding $k$-resolved weights are shown in Figs.~\ref{weight_snap}(e)--(h). Here, the Bloch states forming the $v^+$- and $v^-$-wave are indicated by the green and orange ovals, respectively in Figs.~\ref{weight_snap}(a) and \ref{weight_snap}(e). For $t\lesssim60$, the $k$-resolved weights are in the RH chain for the all cases, and thus the weights in the LH chain are 0 [Figs.~\ref{weight_snap}(b), \ref{weight_snap}(d), \ref{weight_snap}(f), and \ref{weight_snap}(h)]. For $t\gtrsim60$, the $v^+$-wave enters the LH chain. One can notice that the Bloch states forming the $v^+$-wave in the LH chain remain the same branch in the $k$-space for $\phi=0$ [Figs.~\ref{weight_snap}(b) and \ref{weight_snap}(f)], while they move toward ``right'' (``left'') for $\sigma_I = \downarrow$ ($\sigma_I=\uparrow$), i.e., the other branch with positive group velocity for $\phi=2\pi/3$ [Figs.~\ref{weight_snap}(d) and \ref{weight_snap}(h)]. Thus, the wavenumber is approximately conserved after passing through the CIF for $\phi=0$, while it is not for $\phi=2\pi/3$. As shown in Figs.~\ref{band_spin}(a) and \ref{band_spin}(b), $S_{z,k,\mu}^R = -S_{z,k,\mu}^L$, which means that the direction of the spin in the $v^+$-wave in the LH chain after the scattering at the CIF is flipped for $\phi=0$. In contrast, $S_{z,k,\mu}^L$ for the branch with the large weight in the LH chain in Figs.~\ref{weight_snap}(d) and \ref{weight_snap}(h) is negative for $\phi=2\pi/3$. This is consistent with the data without spin flip for $\phi=2\pi/3$ in Fig.~\ref{Sz_snap}(b). 

\begin{figure}[t!h]
\centering
\includegraphics[width=1.0\linewidth]{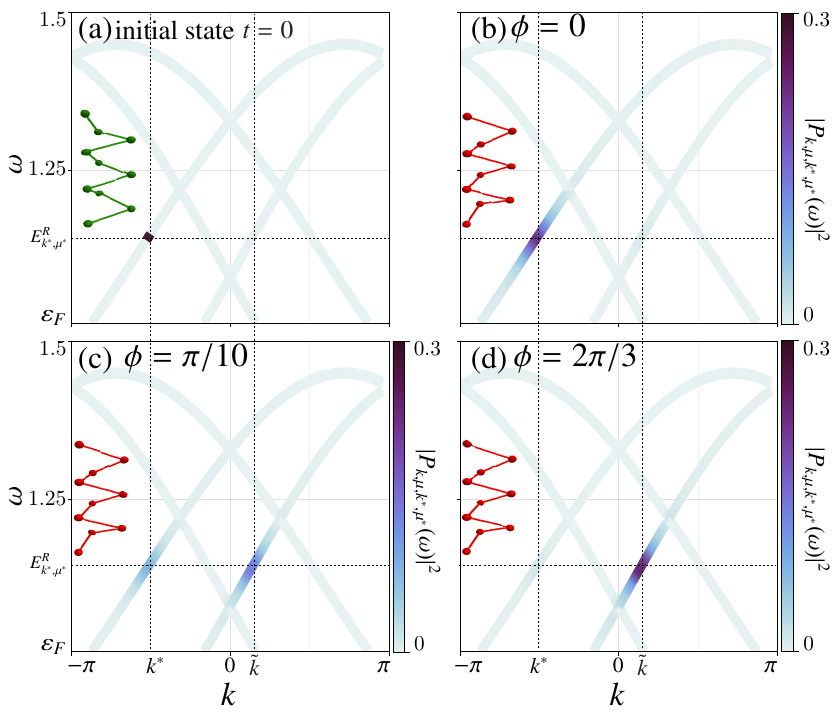}
\caption{(Color online) The scattering probability $|P_{k,\mu,k^*,\mu^*}(\omega)|^2$ for the initial state $\ket{R;k^*,\mu^*}$, where $k^* =\pi/2$ and $\mu^* = 16$. (a) The initial state $\ket{R;k^*,\mu^*}$. Since this state is in the right-handed chain, $S_z$ and $M_z$ take negative values according to Figs.~\ref{band_spin}(a) and \ref{band_spin}(c). (b)-(d) $|P_{k,\mu,k^*,\mu^*}(\omega)|^2$ for  (b) $\phi=0$, (c) $\phi=\pi/10$, and (d) $\phi=2\pi/3$.}
\label{transition}
\end{figure}

As for the orbital angular momentum and the magnetization, we can carry out the similar argument. The quantitative difference is that the both bands with the positive group velocity possess the negative (positive) orbital angular momentum in the RH (LH) chain [See Figs.~\ref{band_spin}(b) and (c)]. Therefore, the orbital angular momentum flips regardless of the angle $\phi$. For the magnetization, some bands possess small $|M_{z,k,\mu}^{R(L)}|$ as shown in Figs.~\ref{band_spin}(c) and \ref{band_spin}(f). As a result, the wave packet with the negative magnetization almost vanishes in the LH chain for both $\phi=0$ [Fig.~\ref{Sz_snap}(c)] and $\phi=2\pi/3$ [Fig.~\ref{Sz_snap}(d)]. This suggests the CIF works as a magnetization filter prohibiting negative $M_z$ inside the LH.

\subsection{The scattering probability of the single Bloch state}

Let us now examine the scattering probability $|P_{k,\mu,k^*,\mu^*}(\omega)|^2$ defined in Eq.~(\ref{trans_prob}). Here, we use $k^*$ and $\mu^*$ instead of $k'$ and $\mu'$ for the wavenumber and the band index of the initial state. 

To address the correspondence between the following analysis and the results in Sec.~\ref{subsec_timeevo}, we choose the initial state with $k^* = -\pi/2$ and $\mu^* = 16$. This is one of the Bloch states constituting the $v^+$-wave in the RH chain for $\sigma_I = \downarrow$ [see Figs.~\ref{weight_snap}(a) and \ref{weight_snap}(c)], as shown in Fig.~\ref{transition}(a). The probability of finding an electron in $\ket{L;k,\mu}$ is shown in Figs.~\ref{transition}(b)--\ref{transition}(d). For $\phi=0$ [Fig.~\ref{transition}(b)], $|P_{k,\mu,k^*,\mu^*}(\omega)|^2$ is finite at $k \approx k^*$ and $\omega \approx E_{k^*,\mu^*}^R$, where the direction of the spin and also the magnetization are opposite in the RH and LH chain. This suggests that the spin flip occurs for $\phi=0$, which is consistent with the result in Sec.~\ref{subsec_timeevo}. However, as $\phi$ increases, a finite $|P_{k,\mu,k^*,\mu^*}(\omega)|^2$ emerges at $k \approx \tilde{k}> 0$ and $\omega \approx E_{k^*,\mu^*}^R$, i.e., the other band with the positive group velocity, where the sign of the spin is the same as the initial state. See Figs.~\ref{transition}(c) and \ref{transition}(d). For $\phi=2\pi/3$ [Fig.~\ref{transition}(d)],  $|P_{k,\mu,k^*,\mu^*}(\omega)|^2$ at $k \approx k^*$ almost vanishes and it is finite only in the other band at $k \approx \tilde{k}$, which implies that the spin does not flip for this case. From these results together with the previous ones in Sec.~\ref{subsec_timeevo}, we can conclude that the spin and the magnetization flip when the momentum and the energy are approximately conserved during the scattering processes in this system. 
\begin{figure}
\centering
\includegraphics[width=1.0\linewidth]{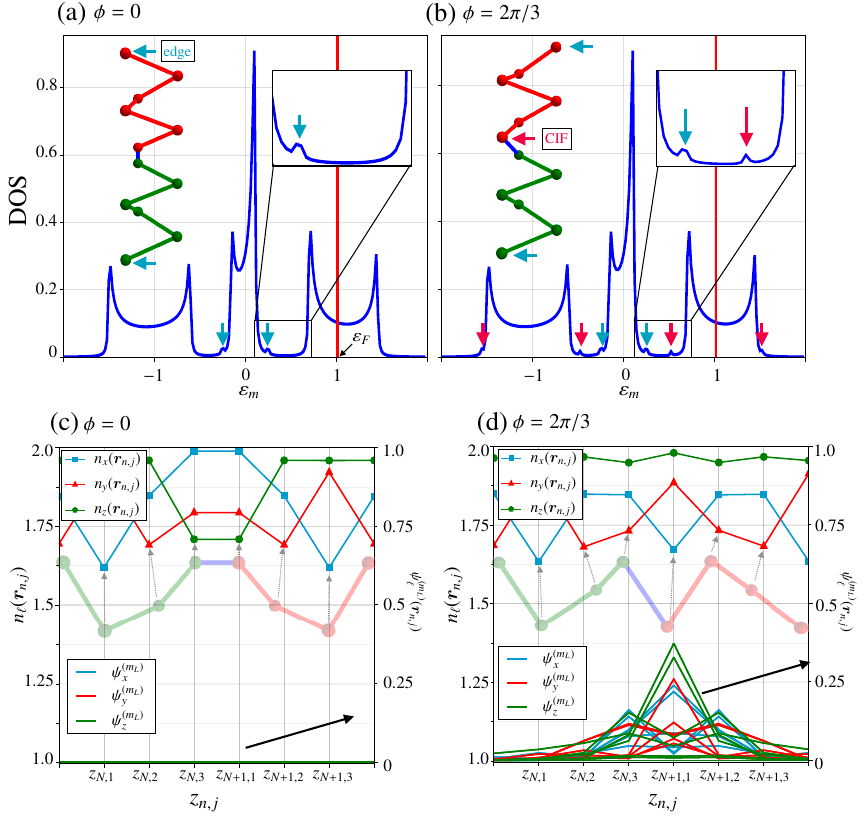}
\caption{(Color online) DOS for (a) $\phi=0$ and (b) $\phi=2\pi/3$. The red (sky blue) arrows indicate the localized state near the CIF (edges of chains). Corresponding positions are schematically shown. The insets show a zoom-in near the energy where the localized states exist. (c) and (d) The number density of electrons in the $p_\ell$ orbital $n_\ell(\bm{r}_{n,j})$ (left axis) and the  amplitude of the wave function of the localized modes $\psi_\ell^{(m_L)}(\bm{r}_{n,j})$ (right axis).}
\label{DOS}
\end{figure}
\subsection{Density of states and localized modes}
To discuss the relation between the electronic state and the spin-flip processes, we investigate the DOS, the local wave functions, and the number density at the CIF. Such localized states at interfaces have been observed along nematic domain walls in the scanning tunneling microscope measurement in FeSe\cite{nematic} and as explained below it is the key to the spin-flip mechanism in this system.

Figures~\ref{DOS}(a) and \ref{DOS}(b) show the DOS for $\phi=0$ and $\phi=2\pi/3$, respectively. Apart from three main separate parts, one can find spikes due to the localized modes at the edges and those at the CIF as indicated by arrows. The modes localized at the edges of the chain are found in both cases. However, the localized modes at the CIF are only found for $\phi=2\pi/3$. Figures~\ref{DOS}(c) and \ref{DOS}(d) show the number density  $n_\ell(\bm{r}_{n,j})\equiv\sum_{\sigma=\uparrow,\downarrow}\ev{\hat{p}^\dagger_{n,j,\ell,\sigma}\hat{p}_{n,j,\ell,\sigma}}{F}$ and the amplitude of the wave function of the localized modes $m=m_L$, $\psi_\ell^{(m_L)}(\bm{r}_{n,j})\equiv\sum_{\sigma}\ev{\hat{p}^\dagger_{n,j,\ell,\sigma}\hat{p}_{n,j,\ell,\sigma}}{m_L}$ for $\phi=0$ and $\phi=2\pi/3$, respectively. This clearly shows there are localized modes at the CIF for $\phi=2\pi/3$  [Fig.~\ref{DOS}(c)] and the $p_z$ component $\psi_z^{(m_L)}$ at the CIF is the largest among the three components followed by $p_y$ and $p_x$ orbital components. Thus, the number density of the $p_z$ orbital $n_z(\bm{r}_{n,i})$ is the largest among the three. In contrast, there are no localized modes for $\phi=0$ [Fig.~\ref{DOS}(c)] and $n_z(\bm{r}_{n,i})$ is the lowest.
\section{Discussion} \label{sec:discussion}
So far, we have demonstrated that the spin flip occurs for $\phi=0$, where the localized modes at the CIF do not exist. In contrast, the spin does not flip when there are localized modes at the CIF for $\phi=2\pi/3$. In the former case, the wavenumber of the electrons passing through the CIF is approximately the same, while in the latter, that is not conserved. For both cases, the energy is approximately conserved. In this section, we will discuss the mechanism of the spin flip at the CIF by analyzing the matrix elements of the SOC and $\phi$ dependence of the total spin for the wave packet in the LH chain. 
\subsection{Mechanism of spin flip at the CIF}\label{sec:dis1}
In our model [Eqs.~(\ref{HR}) and (\ref{V_inter})], the SOC accounts for the spin-flip scattering at the CIF. We first analyze the matrix elements of the SOC defined in Eq.~(\ref{SOC_mat}) in detail. One of the $2\times2$ off-diagonal block matrix between $\lbrace p_{x\uparrow},p_{x,\downarrow}\rbrace\equiv\lbrace p_x\rbrace$ and $\lbrace p_{y\uparrow},p_{y,\downarrow}\rbrace\equiv\lbrace p_y\rbrace$ has a form
\begin{align}
\begin{pmatrix}
i&0\\
0&-i
\end{pmatrix},
\end{align}
which is obviously diagonal in the spin index. However, the block matrices containing $\lbrace p_{z\uparrow},p_{z,\downarrow}\rbrace\equiv\lbrace p_z\rbrace$ such as $\lbrace p_x\rbrace\text{-}\lbrace p_z\rbrace$ and $\lbrace p_y\rbrace\text{-}\lbrace p_z\rbrace$ are 
\begin{align}
\begin{pmatrix}
0&1\\
-1&0
\end{pmatrix}\quad {\rm and}\quad
\begin{pmatrix}
0&i\\
i&0
\end{pmatrix},
\end{align}
respectively. Note that they have off diagonal elements in the spin index. Thus, any spin-flip processes can occur only through the transitions to and from the $p_z$ orbital. 

\begin{figure}
\centering
\includegraphics[width=1.0\linewidth]{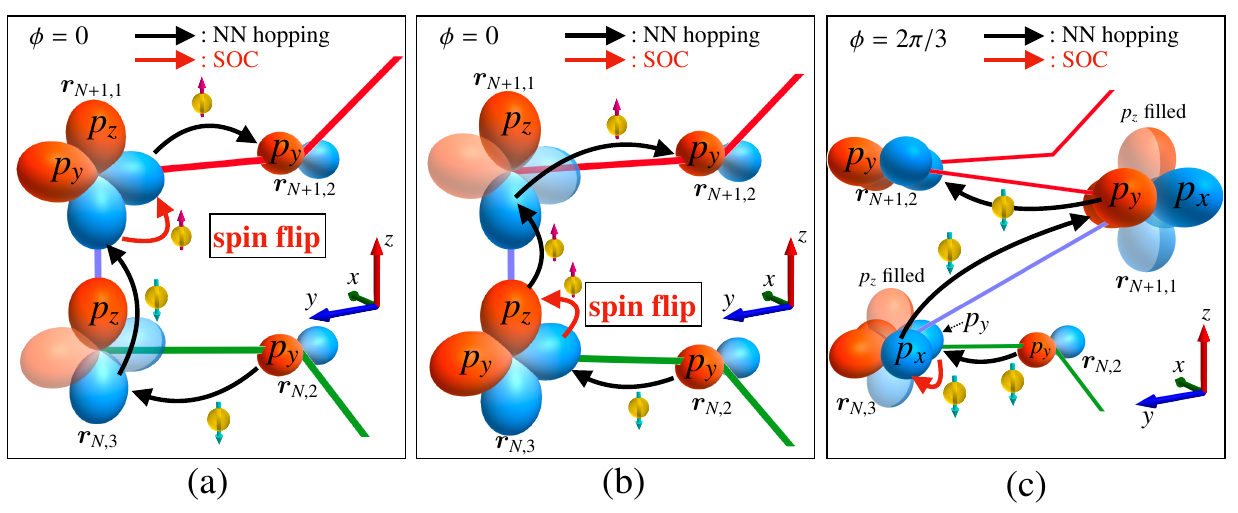}
\caption{(Color online) Schematic pictures of spin-flip and no spin flip processes in the first order of $\lambda$. (a) The $p_z$-orbital electron at $\bm{r}_{N,3}$ hops to the $p_z$ orbital at $\bm{r}_{N+1,1}$ and then the spin is flipped by the SOC for $\phi=0$. (b) The  $p_y$-orbital electron at $\bm{r}_{N,2}$ hops to the $p_y$ orbital at $\bm{r}_{N,3}$. The spin is flipped at the same site $\bm{r}_{N,3}$ with the orbital changed into $p_z$ via the SOC for $\phi=0$. Then, the $p_z$ electron can hop to the LH chain. (c) For $\phi=2\pi/3$, the spin of $p_y$ orbital electron is not flipped by the SOC since the $p_z$ orbital is almost occupied. Thus, the SOC leads to the orbital flip from $p_y$ to $p_x$ without the spin flip. Then the $p_x$ electron hops to the LH chain.}
\label{spin_flip_mechanism}
\end{figure}

Now, consider an electron which propagates in the RH chain for $z<0$ toward the CIF and hops from $\bm{r}_{N,2}$ to $\bm{r}_{N,3}$. Figures~\ref{spin_flip_mechanism}(a) and \ref{spin_flip_mechanism}(b) illustrate the schematic spin flip processes for $\phi=0$. The incoming electron from $\bm{r}_{N,2}$ hops to either the $p_z$ or $p_y$ orbital at $\bm{r}_{N,3}$ because the bond $\bm{d}^N_{2,3}$ between $\bm{r}_{N,2}$ and $\bm{r}_{N,3}$ only has the $y$ and $z$ components and the hopping is constrained to the $\sigma$-bond. Let us explain possible scattering processes with the spin flip in the first order in the SOC. When the electron hops to the $p_z$ orbital at $\bm{r}_{N,3}$ as shown in Fig.~\ref{spin_flip_mechanism}(a), it can directly hop to the $p_z$ orbital at $\bm{r}_{N+1,1}$ since the bond is along the $z$ direction. Then, the electron hybridizes to the $p_y$-orbital with the spin flipped due to SOC and it hops to $\bm{r}_{N+1,1}$. In the case shown in Fig.~\ref{spin_flip_mechanism}(b), the electron in the $p_y$-orbital hybridizes to the $p_z$-orbital through the SOC at $\bm{r}_{N,3}$ with the spin flipped and hops toward the LH chain for $z>0$. In clear contrast, for $\phi=2\pi/3$, the $p_z$ orbital is almost occupied as shown in Fig.~\ref{DOS}(d), where the $p_z$-dominant localized modes at the CIF are the key to this. This suggests that the incoming electron cannot enter the $p_z$ orbital at $\bm{r}_{N,3}$ from $\bm{r}_{N,2}$. This also means that the $p_y$ electron hops to either the $p_x$ or $p_y$ orbital at $\bm{r}_{N+1,1}$, since the $p_z$ orbital is almost occupied at $\bm{r}_{N+1,1}$. During the hopping process above, one can consider the effects of the SOC. However, $\lbrace p_x\rbrace\text{-}\lbrace p_y\rbrace$ block of SOC is diagonal in the spin index, and thus, the spin is not flipped. These arguments qualitatively explain the numerical results in Sec. 4, which show that the orbital degrees of freedom together with the SOC near the CIF play an  important role in the spin-flip scattering at the CIF.

\subsection{$\phi$ dependence}
\begin{figure}
\centering
\includegraphics[width=1.0\linewidth]{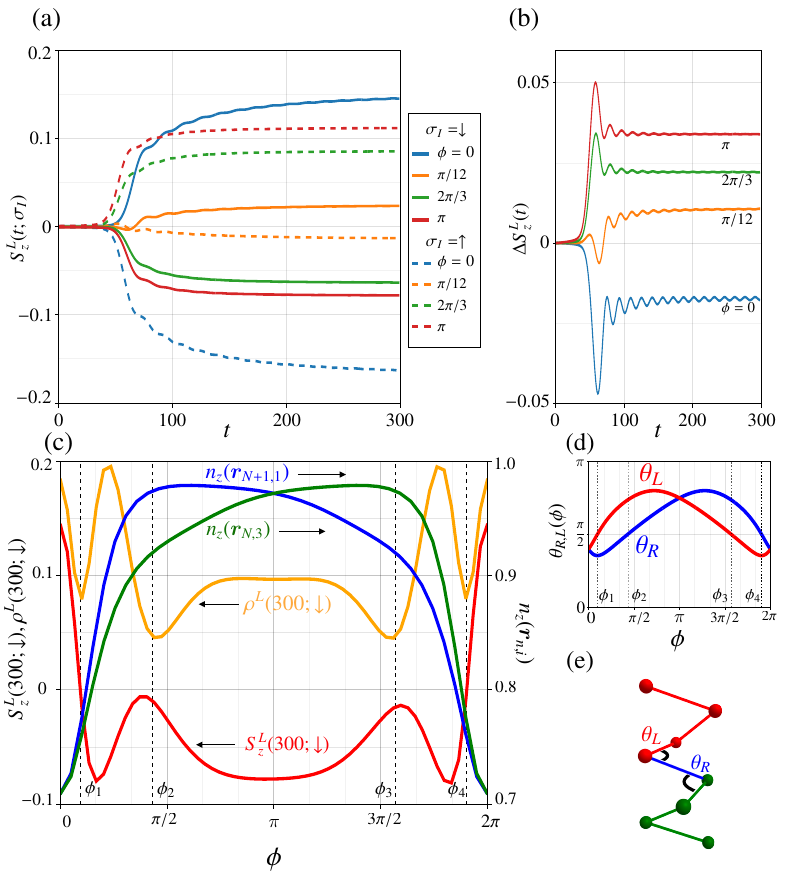}
\caption{(Color online) (a) Total spin angular momentum in the LH chain $S_z^L(t;\sigma_I)$ for several $\phi$'s for the initial spin $\sigma_I=\downarrow$ (solid line) and $\sigma_I=\uparrow$ (dashed line). (b) $\Delta S^L(t)$ for several $\phi$'s. (c) $\phi$ dependence of $S_z^L(t=300;\downarrow)$ (red), $\rho^L(t=300;\downarrow)$ (orange), $\bm{r}_{N,3}$ (green), and $\bm{r}_{N+1,1}$ (blue). (d) $\phi$ dependence of $\theta_R$ (blue) and $\theta_L$ (red). (e) Schematic picture of angles $\theta_R$ and $\theta_L$.}
\label{angle_dependence}
\end{figure}

So far, we have focused on the spin-flip processes in the two cases, $\phi=0$ and $\phi=2\pi/3$. In order to check the validity of the analyses in Sec. 5.1, we examine the $\phi$ dependence. We define the total $S_z(\bm{r}_{n,j},t)$ in the LH chain as $S_z^L(t;\sigma_I)\equiv\sum_{n,i}^{N+1\leq n \leq 2N} S_z(\bm{r}_{n,j},t)$ with the initial spin $\sigma_I$. Figure~\ref{angle_dependence}(a) shows $S_z^L(t;\sigma_I)$ for several values of $\phi$ for $\sigma_I=\downarrow$ (solid line) and $\sigma_I = \uparrow$ (dashed line). For all $\phi$'s, $S_z^L(t; \sigma_I)=0$ for $t\lesssim50$ since the wave packet is in the RH chain. For $t\gtrsim50$, $S_z^L(t; \sigma_I)$ is finite and becomes nearly constant at $t\approx300$, which means that the wave packets no longer enter the LH chain. $S_z^L(t; \downarrow)>0$ and $S_z^L(t; \uparrow)<0$ indicate that the spin is flipped.  Although the $\phi$ dependence for the two cases $\sigma_I=\uparrow$ and $\downarrow$ are qualitatively similar when comparing $S_z^L(t; \downarrow)$ and $-S_z^L(t; \uparrow)$, they are slightly different. Figure~\ref{angle_dependence}(b) shows $ \Delta S^L(t) \equiv S_z^L(t,\downarrow)+S_z^L(t;\uparrow)$. As expected from the difference in the electronic states for $\sigma_I=\uparrow$ and $\downarrow$, $\Delta S^L(t)$ is finite and the absolute value is approximately $\sim$10 \% of $|S_z(t;\sigma_I)|$. This suggests that a finite spin polarization is generated in the LH chain even when the injected electron is unpolarized and the polarization depends on the angle $\phi$. Note that $|\Delta S_z(t)|>0$ is in general the case but the results can be quantitatively modified when taking into account the interference between the $\sigma_I=\uparrow$ and $\downarrow$ wavepackets.

Now, we discuss the $\sigma_I = \downarrow$ case in detail since qualitatively the same results are obtained for $\sigma_I=\uparrow$. In Fig.~\ref{angle_dependence}(c), we show the $\phi$ dependence of $S_z^L(300;\downarrow)$ and the total number of electrons in the LH chain $\rho^L(t;\downarrow)\equiv\sum_{n,j}^{N+1\leq n \leq 2N} \rho(\bm{r}_{n,j},t)$, where $\rho(\bm{r}_{n,j},t) = \ev{\sum_{\ell,\sigma}\hat{p}^\dagger_{n,j,\ell,\sigma}(t)\hat{p}_{n,j,\ell,\sigma}(t)}{I}$.
As $\phi$ increases, $S_z^L(300;\downarrow)$ rapidly decreases and becomes negative, which means that the spin does not flip at the CIF. Meanwhile, the density of $p_z$ electron at the CIF, $n_z(\bm{r}_{N,3})$ and $n_z(\bm{r}_{N+1,1})$ increase nearly as rapid as $S_z^L(300;\downarrow)$ as shown in Fig.~\ref{angle_dependence}(c). This evidences that the localized $p_z$ electrons at the CIF prevent the incoming electrons from entering the $p_z$ orbital and it suppresses the spin-flip scattering at the CIF. The increase in $S_z^L$ for $\phi\gtrsim \pi/6$, which seems contradicting to our analyses in Sec.~\ref{sec:results}, is due to the reduction of the number of electrons entering the LH chain. Near the $\phi$'s where $S_z^L$ takes the local maximum, $\rho^L(300;\downarrow)$ does the local minimum as shown in Fig.~\ref{angle_dependence}(c). This means that there are fewer electrons in the LH chain, and thus, the absolute value of the spin decreases. There are also several $\phi$'s where $\rho^L(300;\downarrow)$ takes the local minimum: $\phi\approx\phi_1 = 0.097\pi, \phi_2 =0.432\pi, \phi_3=1.568\pi, \phi_4=1.903\pi$. At these values of $\phi$'s, the inter-chain bond $\bm{d}_{\text{CIF}}$ becomes orthogonal to either $\bm{d}_{2,3}^N$ or $\bm{d}_{1,2}^{N+1}$, which leads to the reduction in $\rho^L(300;\downarrow)$. 

Figure~\ref{angle_dependence}(d) shows the $\phi$ dependence of the angle $\theta_R$ between $\bm{d}_{\text{CIF}}$ and $\bm{d}_{2,3}^N$ and $\theta_L$ between $\bm{d}_{\text{CIF}}$ and $\bm{d}_{1,2}^{N+1}$. See Fig.~\ref{angle_dependence}(e) for the schematic picture of the angles $\theta_R$ and $\theta_L$. It is obvious that either $\theta_R$ or $\theta_L$ becomes $\pi/2$ at $\phi_\alpha$ ($\alpha=1,2,3,4$). The $\phi$ dependence $\theta_{L,R}(\phi)$ is symmetric with respect to $\phi=\pi$, i.e., $\theta_R(\phi) = \theta_L(2\pi-\phi)$. It is natural that this interchange does not affect the number of electron in the LH chain at the sufficiently long time and thus $\rho^L(300;\downarrow)$ is symmetric with respect to $\phi=\pi$. However, the orbital angular momentum of the electrons is not necessarily the same for $\phi$ and $2\pi-\phi$ and this also affects the spin angular momentum through the SOC. This is why  $S_z^L(300;\downarrow)$ is not symmetric with respect to $\phi=\pi$ in Fig.~\ref{angle_dependence}(c).

The $\phi$ dependence of $S_z^L(t)$ clearly indicates the importance of the orbital profile of electronic states near the CIF as discussed in Sec.~\ref{sec:dis1}. This also means that the design of the interface between the two chains with opposite handedness can influence the spin transport properties through the CIF. For the reference, we also examine scattering processes at an interface between the RH chain and a simple one-dimensional chain without SOC in Appendix B. 
For this case, the spin flip is not observed. Although some electrons flip their spin at the interface, this rate is smaller than that with spin-conserving processes. Furthermore, most of electrons are reflected back at the interface since the wave functions of the two chains are completely different.  This result emphasizes the importance of the CIF for spin transport properties. 

In this study, the control parameter for the structure of the interface is only $\phi$. There are many other candidate parameters. For example, instead of directly connecting the two chains, one can introduce intermediate atoms between them and the electronic configuration of the atoms also affects the physics. Exploring such systems is an interesting problem and we leave the analyses about such setup as our future studies.

 \section{Conclusion} \label{sec:conclusion}
 We have analyzed electronic states near the CIF and the spin transport through the CIF. Our setup is a one-dimensional $p$-electron model with the SOC and consisting of RH and LH chains connected at the CIF. We have carried out analyses of spin-polarized wave packet dynamics, one-electron scattering process, and the electronic states near the CIF. Our microscopic analysis reveals that the spin-flip scatterings at the CIF strongly depend on the local electronic states at the CIF. The geometrical aspects related to the orbital profile of the localized states together with the spin-flip processes due to the SOC play an important role in determining whether the electron spin flips at the CIF or not. In particular, the key in our system is the $p_z$ orbital at the CIF. This opens a way to manipulate the direction or polarization of the spin currents at the CIF by controlling the localized electron states without magnetic fields. For the deeper understanding and other mechanisms of the spin-flip processes in different systems further studies are needed and we leave them as our future problems.



\section*{Acknowledgement(s)}
The authors thank Takayuki Ishitobi, Takuya Nomoto, Takashi Hotta, Rikuto Oiwa, and Hiroaki Kusunose for fruitful discussions. This work was supported by a Grant-in-Aid for Transformative Research Areas (A) ``Asymmetric Quantum Matters'', JSPS KAKENHI Grant Number JP23H04869.


\appendix
\section{The Effect of $\pi$-bond Hopping}
In this Appendix, we consider the effects of $\pi$-bond hoppings on the spin transport at the CIF discussed in the main text. Since the symmetry around the NN bond is low, there are terms which cannot be described by the Slater-Koster parameters. We here analyze the effects of the $\pi$-bonds as a representative components among subleading hopping processes.
\begin{figure}
\centering
\includegraphics[width=1.0\linewidth]{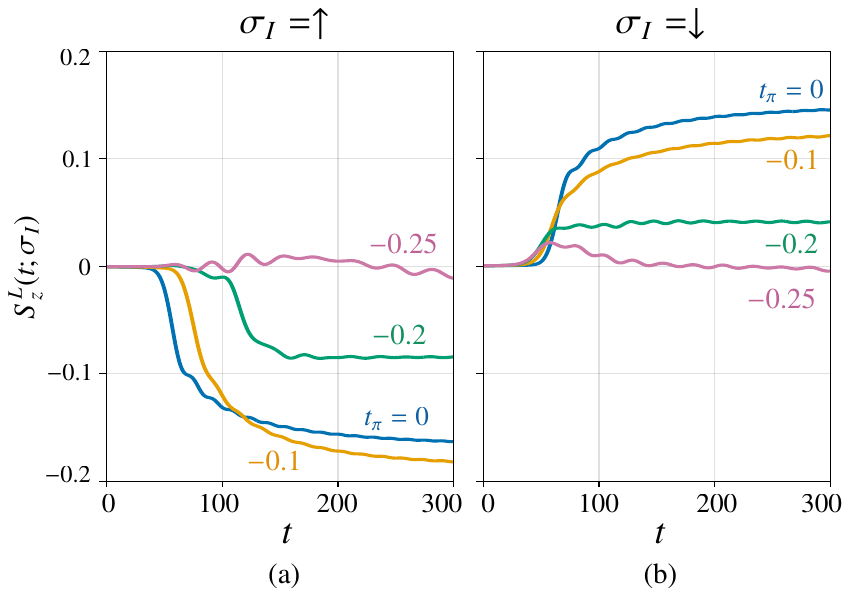}
\caption{(Color online) Spin moment in the LH chain $S_z^L(t)$ as a function of $t$ for $t_\pi = $ 0, $-0.1$, $-0.2$, and $-0.25$ with $\phi = 0$. (a) $\sigma_I = \uparrow$ and  (b) $\sigma_I = \downarrow$.}
\label{pi_hop}
\end{figure}

The Hamiltonian for the $\pi$-bond hopping term can be described as
\begin{align}
H_{\pi}= \ t_{\pi}\sum_{n,i,\sigma}^{\substack{1\leq n \leq N \\(N+1\leq n\leq 2N)}}&\Big[(\hat{\bm{p}}^\dagger_{n,j+1,\sigma}\cdot\bm{u}^n_{j,j+1})(\hat{\bm{p}}_{n,j,\sigma}\cdot\bm{u}^n_{j,j+1})\notag\\
&+(\hat{\bm{p}}^\dagger_{n,j+1,\sigma}\cdot\bm{v}^n_{j,j+1})(\hat{\bm{p}}_{n,j,\sigma}\cdot\bm{v}^n_{j,j+1})\Big]+\text{h.c.},
\end{align}
where $t_{\pi}<0$ is the $\pi$-hopping parameter and $\bm{u}_{j,j+1}^n$ and $\bm{v}_{j,j+1}^n$ are the direction of two $\pi$-bonds which are perpendicular to $\bm{d}_{j,j+1}^n$. As mentioned in Sec.~2.1, contribution of $\pi$-bonds is smaller than that of $\sigma$-bond. Therefore, we examine the spin flip processes when $t_\pi$ is smaller than the value of the $\sigma$-bond hopping set to 1.

Figure~\ref{pi_hop} shows the total spin $S_z^L(t)$ introduced in Sec. 5.2 for  $t_\pi=0, -0.1, -0.2$ and $-0.25$: (a) $\sigma_I = \uparrow$ and (b) $\sigma_I = \downarrow$. We have set $\phi = 0$ in this analysis. It should be noted here that electrons can hop from the RH chain to the LH chain via the $\pi$-bond hoppings without through the intermediate $p_z$-orbital at the interface. Therefore, $|S_z^L(t)|$ tends to be suppressed as increasing $|t_\pi|$. Although there is quantitative suppression in $|S_z^L(t)|$ due to the finite $t_{\pi}$, our results in Secs.~4 and 5 are qualitatively still valid for moderate magnitude of $t_{\pi}$.

\section{An Interface Between the Chiral Chain and a Non-Chiral Chain}
\begin{figure}
\centering
\includegraphics[width=1.0\linewidth]{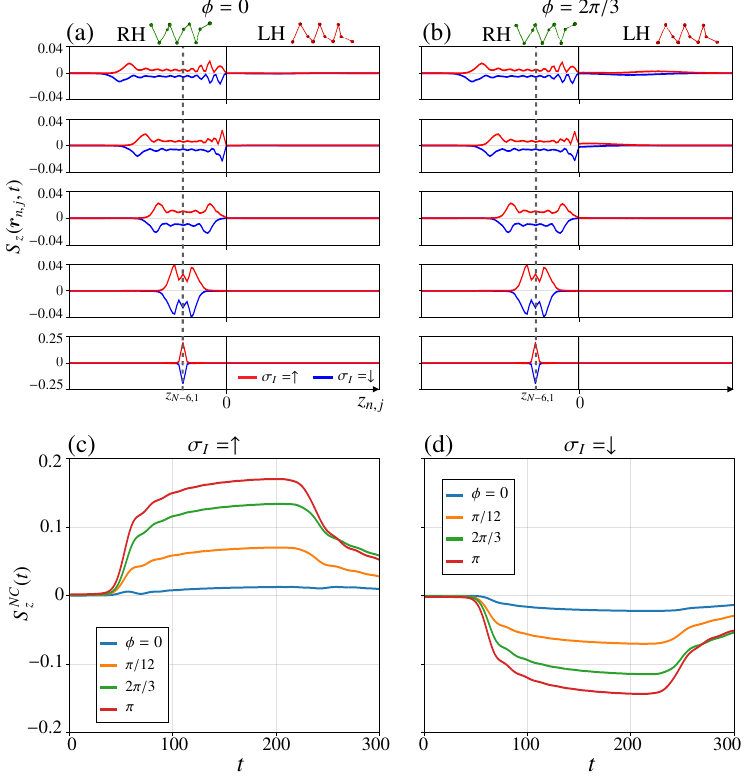}
\caption{(Color online) $S_z(\bm{r},t)$ for $\sigma_I=\uparrow$ (blue) and $\sigma_I=\downarrow$ (red) as a function of $\bm{r}$, where $\bm{r} = \bm{r}_{n,j}$ for $1\leq n \leq N$ and $\bm{r} = \tilde{\bm{r}}_{n}$ for $N+1\leq n \leq 4N$, for (a) $\phi=0$ and (b) $\phi=2\pi/3$. 
 $S_z^{NC}(t)$ as a function of $t$ for $\phi=0,\pi/12, 2\pi/3$, and $\pi$ for (c)  $\sigma_I = \uparrow$ and (d) $\sigma_I = \downarrow$.}\label{nonchiral}
\end{figure}

In this Appendix, we consider the situation where there is an  interface between the chiral chain and one-dimensional non-chiral (NC) chain. The latter one-dimensional chain is supposed to consist of $s$-orbital electrons and their tight-binding Hamiltonian is given as
\begin{align}
H_{NC} = \sum_{n,\sigma}^{N+1\leq n \leq 4N-1}\hat{s}^\dagger_{n+1,\sigma}\hat{s}_{n,\sigma} + \text{h.c.},
\end{align}
where $\hat{s}^\dagger_{n,\sigma}$ and $\hat{s}_{n,\sigma}$ are the creation and annihilation operators of the $s$-electron with the spin $\sigma$ at the $n$th unit cell. The total number of unit cells in the NC chain is $3N$ so that both the RH chain and the NC chain have the same number of sites. We denote the position of the $n$th unit cell in the NC chain as $\tilde{\bm{r}}_n$ and set $\tilde{\bm{r}}_{N+1} = \bm{r}_{N+1,1}$. Then, the full Hamiltonian is obtained by replacing $H_L$ [Eq.~(\ref{HR})] with $H_{NC}$ in $H = H_R + H_L + V$, where now $V$ [Eq.~(\ref{V_inter})] is replaced by
\begin{align}
V = \sum_{\sigma}\hat{s}^\dagger_{N+1,\sigma}(\bm{p}_{N,3,\sigma}\cdot\bm{d}_{\text{CIF}}) + \text{h.c.},
\end{align}
 We can carry out a similar analysis about this model as we have done in Sec.~\ref{subsec_timeevo}.

Figures~\ref{nonchiral}(a) and (b) show $S_z(\bm{r},t)$ at $t =$ 0, 20, 40, 60, and 80 for $\phi =0$ and $\phi=2\pi/3$, where $\bm{r} =\bm{r}_{n,j}$ for $1\leq n \leq N$ (in the RH chain) and $\bm{r} = \tilde{\bm{r}}_{n}$ for $N+1\leq n \leq 4N$ (in the NC chain). 
As one can clearly see, the spin flip does not occur even for $\phi=0$. Figures \ref{nonchiral}(c) and (d) illustrate the total spin $S_z^{NC}(t) = \sum_{n}^{N\leq n\leq 4N} S_z(\tilde{\bm{r}}_n,t)$ in the NC chain for several values of $\phi$'s for $\sigma_I = \uparrow$ and $\sigma_I = \downarrow$. For any $\phi$, the spin flip is not observed. This can be understood as follows. First, the wave functions of the RH chain and the NC chain are completely different due to the structural difference in these two chains. Therefore, approximately 49\% (47\%) of electrons are reflected at the interface for $\sigma_I=\downarrow$ ($\sigma_I = \uparrow$) even for $\phi=0$, which is 222\% (247\%) of the reflection at the CIF. Here, the reflection rate for the non-CIF chain $R_{NC}$ and CIF chain $R_C$ is calculated as $R_{NC} = \frac{\rho^{\text{in}}_{NC}-\rho_{NC}^{\text{trans}}}{\rho_{NC}^{\text{in}}}$ and $R_{NC} = \frac{\rho^{\text{in}}_{C}-\rho_{C}^{\text{trans}}}{\rho_{C}^{\text{in}}}$, respectively, where $\rho^{\text{in}}_{NC(C)}$ is the total number of electrons in the right-ward wave packet in the RH chain defined as $\rho^{\text{in}}_{NC(C)} = \sum_{n,j,\ell,\sigma}^{n_I\leq n \leq N}\rho(\bm{r}_{n,j},\Delta t)$ and $\rho^{\text{trans}}_{NC(C)}$ is that of electrons in the NC (LH) chain defined as $\rho^{\text{trans}}_{NC} = \sum_{n}^{N+1\leq n\leq 4N}\rho(\tilde{\bm{r}},t=200)$ [$\rho^{\text{trans}}_{C} = \rho^L(200)$]. We choose $\Delta t = 10$ to make sure that the right-ward and left-ward waves are well separated and the right-ward wave stays in the RH chain. 
The remaining electrons can go into the NC chain and the spin flip probability depends on the $\phi$ dependent occupancy of the $p_z$ orbital at the interface. However, as evidenced in Figs.~\ref{nonchiral}(c) and \ref{nonchiral}(d), the sign of the total spin in the NC chain is the same as the initial spin $\sigma_I$ in the RH chain. In addition, some of the processes such as the one illustrated in Fig.~\ref{spin_flip_mechanism}(a) do not occur because there is only $s$ orbital at $\tilde{\bm{r}}_{N+1} = \bm{r}_{N+1,1}$ and the SOC is absent there. As a result, the probability of the spin flip at the interface is further reduced.

This result emphasizes the importance of the CIF. The RH and the LH chains are the mirror image of each other and this structural similarity leads to the similarity in their wave functions. As a result, the reflection rate at the CIF is reduced and more electrons can pass through the CIF.

\end{document}